\documentclass[10pt,letterpaper,twocolumn]{article}

\usepackage{ol2}
\usepackage[draft]{hyperref}
\usepackage{textcomp, amsmath}
\usepackage{graphicx}

\begin{document}

\twocolumn[

\title{Dispersion-based control of modal characteristics for parametric down-conversion in a multimode waveguide}

\author{Micha{\l} Karpi\'{n}ski$^*$, Czes{\l}aw Radzewicz, Konrad Banaszek}
\address{Faculty of Physics, University of Warsaw, ul.\ Ho\.z{a} 69, 00-681 Warszawa, Poland\\
$^*$Corresponding author: mkarp@fuw.edu.pl}

\begin{abstract}
We report generation of near-infrared photon pairs in fundamental spatial modes via type-II spontaneous parametric down-conversion in a periodically poled potassium titanyl phosphate
(KTiOPO$_4$) nonlinear waveguide supporting multiple transverse modes. This demonstrates experimentally a versatile scheme for controlling the spatial characteristics of the produced nonclassical light based on exploitation of intermodal dispersion. The down-converted photons are characterized by the measurement of the beam quality factors in the heralded regime.
\end{abstract}

]

Nonlinear processes in fibers and waveguides are a promising source of nonclassical radiation  \cite{FiberSources1,FiberSources2,FiberSources3,FiberSources4,
PPKTP1,PPKTP2,MoslChrisPRL09,BanaURenOL01} required to realize quantum protocols for communication \cite{QCrypto}, and other emerging applications \cite{OBriFuruNPH09}. Spatial confinement of optical fields enhances the strength of a nonlinear interaction and alters dramatically the spatial characteristics of the produced light, offering new possibilities to engineer its properties. Controlling spatial modes is necessary to achieve high-visibility  nonclassical interference \cite{URenBanaQIC03}, ensures compatibility with fiber optics networks and integrated optical circuits \cite{PoliMattIEEE09} and enables one to exploit spatial encoding of quantum information \cite{SpatialEntanglement1, SpatialEntanglement2}.

In this Letter we demonstrate experimental control of the modal characteristics of non-classical light generated in a multimode waveguide achieved by a careful exploitation of the dispersive properties of this  structure. Specifically, we realized the spontaneous parametric down-conversion (SPDC) process in a periodically poled potassium titanyl phosphate (PPKTP) waveguide exhibiting $\chi^{(2)}$ nonlinearity to produce photon pairs in pure spatial modes despite the multimode nature of the nonlinear structure. We were able to define the spatial profiles for the generated photons by utilizing differences in propagation constants for individual transverse modes, implementing the proposal outlined in \cite{BanaURenOL01,KarpRadzAPL99}. The modal purity was confirmed by a photon counting measurement of the $M^2$ beam quality factors for the generated radiation. This result demonstrates the feasibility of engineering the modal structure of nonclassical radiation in nonlinear waveguides. In particular, it provides a method to generate single-mode radiation in PPKTP waveguides that exhibit a number of attractive features \cite{PPKTP1,PPKTP2,MoslChrisPRL09,BanaURenOL01}, but are multimode at visible wavelengths due to manufacturing limitations. More generally, our approach paves the way towards preparation of more complex forms of spatial entanglement in nonlinear wave guiding structures \cite{MoslChrisPRL09}.

\begin{figure}
\includegraphics{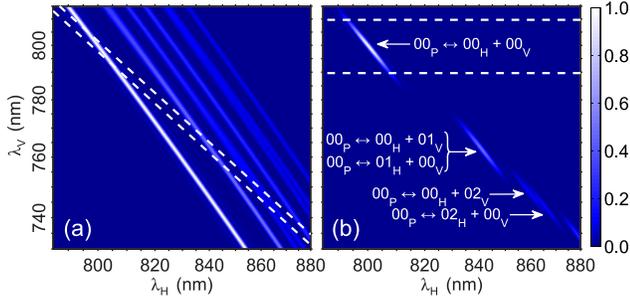}
\caption{(a) A calculated set of phase matching bands corresponding to different pairs of $H$ and $V$ modes for the used PPKTP waveguide with the pump field mode fixed to $00_P$. The dashed lines bound the energy conservation region defined by the pump bandwidth. (b) The resulting joint spectrum of down-converted photon pairs. Horizontal dashed lines represent coarse spectral filtering of $V$ photons which produces heralded $H$ photons in a pure spatial mode. 
The role of $V$ and $H$ photons can be reversed. The axes are labelled with vacuum wavelengths $\lambda_H, \lambda_V$ of the generated photons and scaled linearly in $|1/\lambda|$.}
\label{Fig:maps}
\end{figure}

The principle of our approach is as follows. The process of interest is type-II SPDC in which a broadband pump photon, labeled with an index $P$, is converted into a pair of orthogonally polarized photons denoted as $H$ and $V$. Due to energy conservation, there are two independent spectral parameters. We choose them to be the wavelengths of the $H$ and $V$ photons. In a multimode waveguide, the phase matching condition can be satisfied within a series of bands as exemplified in Fig.~\ref{Fig:maps}(a). Each band, whose spectral width is inversely proportional to the waveguide length, corresponds to a different triplet of spatial modes of the  $P$, $H$, and $V$ photons involved in the nonlinear interaction. We will label the individual modes with a pair of integers $ij$ specifying the number of nodes in two orthogonal directions. The separation of the bands is a consequence of intermodal dispersion, which results in mode-dependent contributions to the phase matching function. Our aim is to generate a pair of photons in the fundamental modes $00_H$ and $00_V$. The best option for the pump beam, maximizing the conversion strength, is the $00_P$ mode \cite{KarpRadzAPL99}. Intermodal dispersion makes the center of the fundamental $00_P \rightarrow 00_V + 00_H$ band shifted from bands involving other combinations of down-converted modes. Further, the spectrum of the pump defines a region within which wavelengths of the down-converted photons satisfy the energy conservation in the SPDC process. Because the photons are generated via a type-II process in a birefringent medium, this region crosses several bands creating a number of islands shown in Fig.~\ref{Fig:maps}(b). For sufficiently narrow pump bandwidth the islands can be separated by coarse spectral filtering of the produced photons, which consequently isolates a single pair of spatial modes, completing our generation scheme.

\begin{figure}
\includegraphics{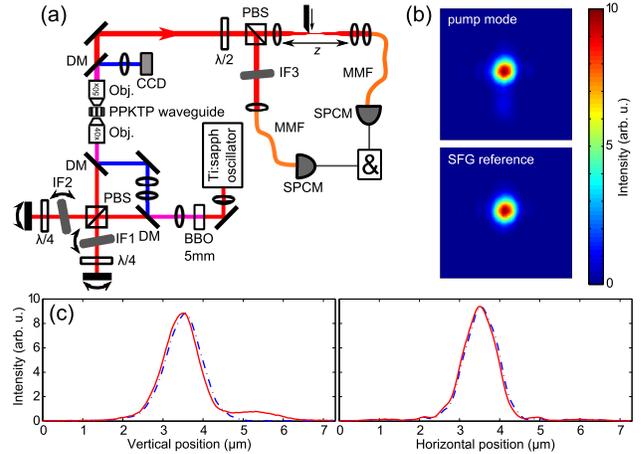}
\caption{(a) The experimental setup. DM, dichroic mirrors; $\lambda/4$, quarter-wave plates; PBS, polarizing beam splitters. For other symbols see the main text. (b) The intensity distribution of the pump field coupled into the waveguide compared with the sum-frequency generated reference. (c) Pump (solid) and SFG reference (dashed) intensity distributions integrated along vertical and horizontal lines.
}
\label{Fig:setup}
\end{figure}

The experimental setup is shown schematically in Fig.~\ref{Fig:setup}(a). At its heart was a temperature stabilized (at $19.0 \pm 0.1^\circ\!\mathrm{C}$) 1~mm long PPKTP structure (AdvR Inc.)  \cite{PPKTP2}, which contained a number of ion-exchanged waveguides designed for efficient type-II second harmonic generation in the $800$ nm wavelength region. In the transverse plane, the waveguides were $2$~$\mu$m wide in the direction parallel to the crystal surface and had an exponentially decaying refractive index profile in the perpendicular direction,  characterized by the effective depth of approximately $5~\mu$m. The mode labels $ij$ refer respectively to the number of nodes in these two directions, and the polarization $H$ is defined as parallel to the crystal surface. Because of the essential role of the phase matching function in our generation scheme, we measured it directly before setting up the down-conversion source. This was done using the mode-resolved sum frequency generation (SFG) spectroscopy technique described in~\cite{KarpRadzAPL99}. The infrared beams were tuned by rotating $0.6$~nm FWHM interference filters IF1 and IF2. For the frequency degenerate process we found the center of the fundamental band at $799.8$~nm with approx.\ $0.7$~nm FWHM bandwidth. The closest band involving $00_P$ mode and higher $H$ and/or $V$ modes was found to be separated by more than $5.0$~nm. Bands corresponding to processes involving higher order $P$ modes were found in the close vicinity of the fundamental band. This meant that selecting the fundamental band relied critically on coupling the pump beam into $00_P$ mode.

To produce photon pairs, a $399.9$~nm pump beam with $1.0$~nm FWHM bandwidth,  chosen to enable spectral separation of the island corresponding to fundamental spatial modes, obtained by frequency doubling $150$~fs pulses from a Ti:sapphire oscillator (Coherent Chameleon Ultra) in a $5$~mm long beta barium borate (BBO) crystal, was coupled into the waveguide. The spatial mode of the beam was matched to the $00_P$ waveguide mode by a combination of lenses, a zoomable beam expander to fine-tune the size of the beam at focus, and a microscope objective. The light exiting the waveguide was collected by a $0.8$ numerical aperture (NA) birefringence-free microscope objective. The pump beam was filtered out by dichroic mirrors and directed to a CCD camera imaging the output face of the waveguide in order to continuously monitor the spatial mode of the pump beam. A comparison of the excited pump beam mode with the reference $00_P$ intensity distribution obtained directly from sum-frequency generation is shown in Fig.~\ref{Fig:setup}(b,c). We used the intensity distributions to calculate the integral of the product of the corresponding normalized mode functions assuming constant phase, obtaining the value of the overlap equal to $0.966 \pm 0.002$.

The spatial properties of the photons produced in the waveguide were characterized by measuring the $M^2$ beam quality parameter via free-space diffraction \cite{ISOM2} at the single photon level.
First, the photons were separated on a polarizing beam splitter. A half-wave plate $\lambda/2$ placed before the polarizer was used to switch the $H$ and $V$ beams between the outputs. The transmitted photons, whose spatial characteristics was to be measured, were focused with a $150$~mm focal length lens to a waist of approximately $50~\mu$m, corresponding to the Rayleigh range $z_R \approx 10$~mm. The transverse distribution was probed with the edge of a knife mounted on a motorized translation stage. The edge could be placed in different locations $z$ in the beam propagation direction by moving it along an optical rail, and swept through the beam in the transverse plane in either horizontal or vertical direction. To comply with the ISO standard of the $M^2$ factor measurement \cite{ISOM2}, we located the knife edge at a minimum of 5 points within the Rayleigh range, and at least $5$ points outside $2z_R$. Light passing the edge was recollimated by another $150$~mm lens, and focused with an $8$~mm focal length aspheric lens (AL) into a $100~\mu$m core diameter multimode fiber (MMF) that delivered the signal to a single photon counting module  (SPCM, Perkin Elmer). We verified that higher-order waveguide modes were coupled into the MMF by exciting them with a macroscopic $800$~nm beam and measuring the transmitted intensity at the MMF output.

\begin{figure}
\includegraphics{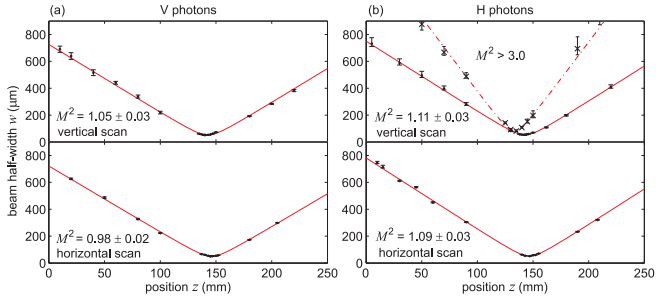}
\caption{The measured beam half-widths $w$ (points with error bars) at positions $z$  for (a) photons $V$ and (b) photons $H$ together with fits of the theoretical dependence (solid lines). The top panel in (b) depicts also results obtained for the pump beam aligned to optimize the coupling efficiency rather than the mode shape (crosses, experimental data; dashed line, fit).}
\label{Fig:m2}
\end{figure}

The essential advantage of the knife-edge method with SPCM detection was the possibility to take measurements in coincidence with the trigger detector monitoring the conjugate photons, used as heralds. The heralds were spectrally filtered with a $10$~nm FWHM interference filter IF3, coupled into a MMF using another 8~mm focal length AL, and delivered to a second SPCM. Signals from both SPCMs were directed to a coincidence circuit with a $6$~ns coincidence window. Approximately $12$~kHz coincidence rate with $15\%$ ratio of coincidence to single count rates in the filtered arm were observed for $45~\mu$W pump power coupled into the waveguide. For every location $z$ and the orientation of the edge, we determined the beam half-width $w$ defined the standard way as doubled
standard deviation for the intensity distribution. In Fig.~\ref{Fig:m2} we depict the dependence $w(z)$ for the $H$ and $V$ photons and the two edge orientations. The experimental points were fitted with the quadratic propagation rule for second-order moments of the intensity distribution \cite{QuadraticPropagation}. The estimated values of the $M^2$ factor, specified in the panels of Fig.~\ref{Fig:m2}, were found to be very close to one.
It is worth stressing that these results were obtained with no spectral filtering of the photons in the scanned arm. When a $10$~nm FWHM spectral filter was inserted also in that arm, we obtained the beam quality factors $M^2 = 1.00 \pm 0.03$ for all the four measurements.

To verify that our measurement setup detected also contributions from higher spatial modes, we changed the coupling of the pump beam into the waveguide to maximize the coupled power rather than to target the excitation of the fundamental pump mode. The resulting pump field intensity distribution was well described by a superposition of $00_P$, $01_P$ and $02_P$ modes and the determination of half-widths yielded values depicted in Fig.~\ref{Fig:m2}(b). Based on these data the beam quality factor can be estimated to exceed $M^2 > 3.0$. This result boldly demonstrates the importance of proper pump beam coupling, in addition to verifying the operation of the setup.

In conclusion, the measurement results presented in this Letter confirm successful generation of photon pairs with a high degree of spatial purity. This demonstrates the feasibility to manage the spatial properties of down-converted photons in a multimode waveguide by combining its dispersive properties with a proper arrangement of the pump field.
In particular, our work provides a direct experimental evidence that the multimode characteristics of extensively studied PPKTP waveguides \cite{PPKTP1, PPKTP2,MoslChrisPRL09,BanaURenOL01} does not compromise their potential to generate photons in single spatial modes compatible with integrated optics circuits and fiber optics transmission. On the other hand, the fact that PPKTP waveguides support multiple spatial modes can be used to generate spatial entanglement \cite{MoslChrisPRL09,KarpRadzAPL99} as well as hyperentangled states \cite{Hyperentanglement}.

We acknowledge insightful discussions with C. Silberhorn and I. A. Walmsley. This work was supported by projects FNP TEAM, FP7 QESSENCE, and the Polish Ministry of Science grant no.\ N~N202~482439.

\end{document}